\documentclass[%
  a4paper, twocolumn, pra, superscriptaddress, amsmath, amssymb, floatfix
]{revtex4-1}
\usepackage{graphicx}
\usepackage{epstopdf}
\usepackage[table]{xcolor}
\usepackage{braket}
\usepackage[colorlinks,bookmarksopen=false]{hyperref}
\usepackage[caption=false]{subfig}
\usepackage{bbm}
\usepackage{lipsum}
\usepackage{siunitx}

\hypersetup{%
  pdfstartview={FitH},
  linkcolor=blue,
  citecolor=blue,
  filecolor=black,
  menucolor=black,
  urlcolor =black
}

\newcommand{\im}{\ensuremath{\textup{i}}}

\newcommand{\ip}{\ensuremath{\mathfrak{Im}}}

\newcommand{\op}[1]{\ensuremath{\mathsf{\hat{#1}}}}
\newcommand{\lop}[1]{\ensuremath{\mathcal{#1}}}

\newcommand{\pulse}{\mathcal{E}}
\newcommand{\dmap}{\mathcal{D}}
\newcommand{\wL}{\omega_{\mathrm{L}}}
\newcommand{\wLz}{\omega_{\mathrm{L}_{0}}}
\newcommand{\wq}{\omega_{\mathrm{q}}}
\newcommand{\wR}{\omega_{\mathrm{R}}}

\newcommand{\dd}{\ensuremath{\mathrm{d}}}

\newcommand{\change}[1]{\textcolor{black}{#1}}

\begin{document}

\title{%
  Reservoir Engineering using Quantum Optimal Control for Qubit Reset
}

\author{Daniel Basilewitsch}
\affiliation{%
  Theoretische Physik, Universit\"{a}t Kassel, D-34132 Kassel, Germany
}

\author{Francesco Cosco}
\altaffiliation[present address: ]{%
  Institute for Theoretical Physics and IQST, Universit\"{a}t Ulm, D-89069 Ulm,
  Germany
}
\affiliation{%
  QTF Centre of Excellence, Turku Centre for Quantum Physics, Department of
  Physics and Astronomy, University of Turku, FI-20014 Turku, Finland
}

\author{Nicolino Lo Gullo}
\affiliation{%
  QTF Centre of Excellence, Turku Centre for Quantum Physics, Department of
  Physics and Astronomy, University of Turku, FI-20014 Turku, Finland
}

\author{Mikko M\"ott\"onen}
\affiliation{%
  QTF Centre of Excellence, Department of Applied Physics, Aalto University,
  FI-00076 Aalto, Finland
}
\affiliation{%
  VTT Technical Research Centre of Finland, P.O. Box 1000, FI-02044 VTT, Finland
}

\author{Tapio Ala-Nissil\"a}
\affiliation{%
  QTF Centre of Excellence, Department of Applied Physics, Aalto University,
  FI-00076 Aalto, Finland
}
\affiliation{%
  Interdisciplinary Centre for Mathematical Modelling and Department of
  Mathematical Sciences, Loughborough University, Loughborough, Leicestershire
  LE11 3TU, United Kingdom
}

\author{Christiane P. Koch}
\affiliation{%
  Theoretische Physik, Universit\"{a}t Kassel, D-34132 Kassel, Germany
}

\author{Sabrina Maniscalco}
\affiliation{%
  QTF Centre of Excellence, Turku Centre for Quantum Physics, Department of
  Physics and Astronomy, University of Turku, FI-20014 Turku, Finland
}
\affiliation{%
  QTF Centre of Excellence, Department of Applied Physics, Aalto University,
  FI-00076 Aalto, Finland
}

\date{\today}

\begin{abstract}
  We determine how to optimally reset a superconducting qubit which interacts
  with a thermal environment in such a way that the coupling strength is
  tunable. Describing the system in terms of a time-local master equation with
  time-dependent decay rates and using quantum optimal control theory, we
  identify temporal shapes of tunable level splittings which maximize the
  efficiency of the reset protocol in terms of duration and error.
  Time-dependent level splittings imply a modification of the system-environment
  coupling, varying the decay rates as well as the Lindblad operators. Our
  approach thus demonstrates efficient reservoir engineering employing quantum
  optimal control. We find the optimized reset strategy to consist in maximizing
  the decay rate from one state and driving non-adiabatic population transfer
  into this strongly decaying state.
\end{abstract}

\maketitle

\section{Introduction} \label{sec:intro}

Superconducting qubits, combining sufficient isolation from the external
environment and good scalability, constitute a promising platform for
demonstrating quantum advantage of a quantum computer~\cite{Gambetta2017}. The
ability to quickly and accurately reset qubits is a key requirement for reaching
the thresholds on state preparation and gate errors required by contemporary
quantum error correction codes. \change{Conventional reset procedures consist of
coupling the qubits to cold environments and waiting for their thermalization.
Although this is effective, it is also slow due to the inherently small coupling
between the qubit and the environment, which sets the time scale of the
thermalization. A faster alternative is to use ancilla systems and to implement
a controlled swap of entropies between the qubit and the
ancilla~\cite{BasilewitschNJP17, PRL.121.060502} or algorithmic
cooling~\cite{PRL.119.050502, NJP.19.113057}. Another alternative is given by}
tunable environments~\cite{Pierre.APL.104.232604, Tan2017, PartanenSciRep18,
Silveri2019, Wong_2019}, which provide a convenient and fast way to initialize
qubits on-demand \change{while still employing the idea of thermalization}.
A method utilizing such a tunable environment to efficiently prepare
superconducting qubits in their ground state has recently been brought
forward~\cite{Tuorila.npjQuantumInf.3.27}. It exploits the indirect coupling of
the qubit to a low-temperature resistive bath via two intermediate
resonators~\cite{Tuorila.npjQuantumInf.3.27} and uses a protocol that utilizes
sequential resonances with the resistive bath. Here, we use quantum optimal
control theory (QOCT) to study the efficiency of this reset protocol.

For a given model of a quantum system and its dynamics, QOCT provides a set of
tools for obtaining the shapes of pulses which maximize a desired objective such
as a gate or state preparation fidelity~\cite{Glaser.EPJD.69.279}. In contrast
to dynamical-decoupling-like approaches~\cite{SuterRMP15}, QOCT does not rely on
any a priori assumptions on the timescales of correlation functions of the
system and the environment, and it allows for continuous dynamical modulation
with minimal restrictions on the shape, duration, and strength of the applied
pulse~\cite{Glaser.EPJD.69.279}. In general, QOCT methods can be distinguished
into those that evaluate only the objective functional such as the chopped
random basis (CRAB) method~\cite{PRA.84.022326} and those that make use of also
the gradient of the objective functional~\cite{Glaser.EPJD.69.279}. The latter
require both forward and backward propagation of the system dynamics and update
the pulse shape either sequentially in time, such as Krotov's
method~\cite{Krotov.book}, or concurrently for all times at once, such as the
gradient ascent pulse engineering (GRAPE) algorithm~\cite{JMagRes.172.296}. In
particular, QOCT is useful to study the control of open quantum systems since it
allows to determine fundamental performance bounds due to decoherence and decay
processes~\cite{KochJPCM16}. Remarkably, the latter are not necessarily
detrimental but may also be desired, for example when export of entropy is
required to reach the objective~\cite{KochJPCM16}. This is true for cooling in
general~\cite{BartanaJCP93, BartanaJCP97, SchmidtPRL11} and especially for reset
of qubits to a pure state~\cite{BasilewitschNJP17, PRA.96.042315, FischerPRA19}.

Utilizing the coupling to environmental degrees of freedom is also at the heart
of quantum reservoir engineering~\cite{PoyatosPRL96} which deliberately
incorporates dissipation into the system dynamics. In its simplest form, it is
realized by a switchable, constant-amplitude electromagnetic field that drives
transitions into a fast decaying state~\cite{PoyatosPRL96}. For open quantum
systems without memory, the system is driven into the fixed point of the
Liouvillian, with constant and positive decay rates, that governs the
dynamics~\cite{KrausPRA08,VerstraeteNatPhys09}. This idea has found widespread
application in quantum optical experiments, for example with trapped
atoms~\cite{KrauterPRL11}, ions~\cite{LinNat13, KienzlerSci14} and circuit QED
platforms~\cite{Doucet2018}. For trapped ions, combining reservoir engineering
with QOCT has recently allowed to determine the field strengths required to
reach the error correction threshold in entangled-state
preparation~\cite{HornNJP18}. For superconducting qubits, major decoherence
arises from two-level fluctuators which also render the dynamics
non-Markovian~\cite{PaladinoRMP14}. This can be captured by a strongly coupled
environmental mode~\cite{RebentrostPRL09} or negative and time-dependent decay
rates in a master equation~\cite{RivasRPP14}. However, reservoir engineering
protocols have thus far been limited to exploiting decay with constant or
piecewise constant rates~\cite{KrausPRA08, ReiterPRA13, MirrahimiNJP14,
KimchiPRL16}.

Here, we lift the limitation of constant decay rates by combining reservoir
engineering with QOCT and a master equation featuring time-dependent rates. The
latter are both controllable and experimentally implementable with current
technologies~\cite{Tuorila.npjQuantumInf.3.27}. Using Krotov's method for
QOCT~\cite{Reich.JCP.136.104103}, we derive the optimal shape of the external
control fields that determine the time-dependent decay rates in the master
equation.

The paper is structured as follows: In Section~\ref{sec:methods}, we introduce
both the Hamiltonian of the model and the main features of the used quantum
optimal control method. In Section~\ref{sec:results}, we present the numerical
results for the optimization of the original protocol and compare it with the
previous solution~\cite{Tuorila.npjQuantumInf.3.27}. Moreover, we extend the
original protocol by adding two additional sets of control fields and evaluate
the influence of the initial fields with which the optimization is started.
Finally, in Section~\ref{sec:concl} we summarize our findings and present the
conclusions of this work.

\section{Model and Methods}\label{sec:methods}

\begin{figure*}[tb]
  \centering
  \includegraphics{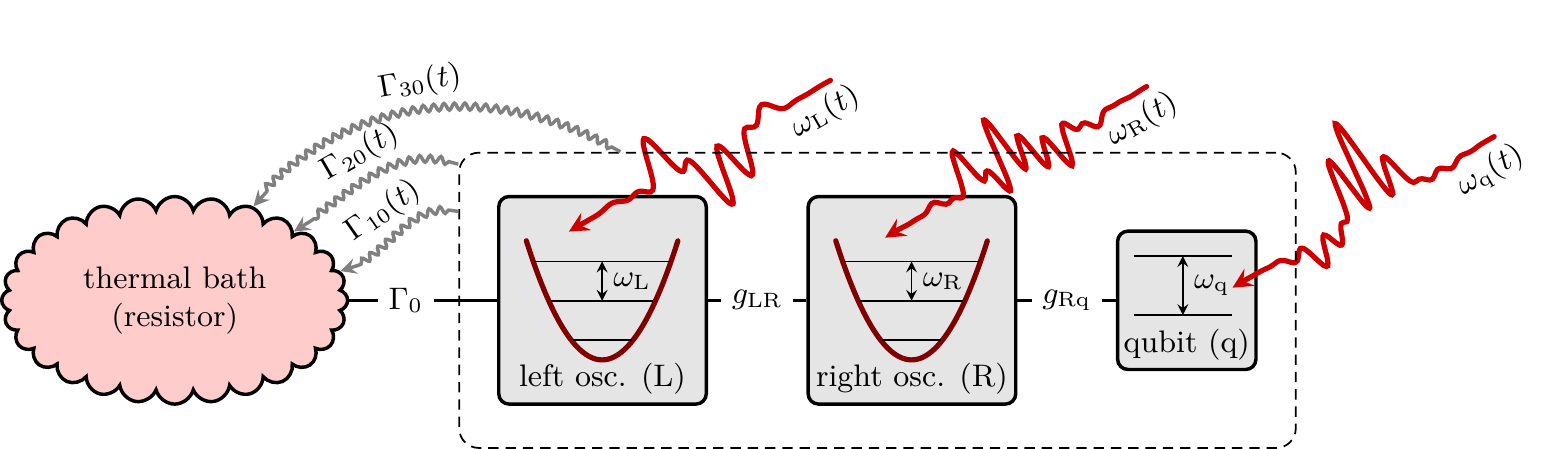}
  \caption{%
    Schematic diagram of the considered physical scenario consisting of a qubit
    (q) linearly coupled to a harmonic oscillator (R), which in turn is linearly
    coupled to a second harmonic oscillator (L) that is in direct contact with
    a thermal bath. By temporally controlling the level splittings
    $\omega_{\mathrm{q/L/R}}(t)$ of the qubit, the right and the left
    oscillator, one can effectively tune the coupling strength to the bath and
    change the decay rates over several orders of magnitude.
  }
  \label{fig:scheme}
\end{figure*}

\subsection{Model}
We consider a three-partite system consisting of two harmonic oscillators, named
left (subscript L) and right (subscript R) oscillator, and a qubit (subscript q)
as sketched in Fig.~\ref{fig:scheme} and previously discussed in
Refs.~\cite{PartanenSciRep18, Tuorila.npjQuantumInf.3.27}. We assume the two
oscillators to be linearly coupled to each other through quadrature operators
and the qubit to be exclusively coupled to the right oscillator. This scenario
is modeled by the Hamiltonian (using units in which $\hbar=1$)
\begin{equation}\label{system-hamiltonian}
  \begin{aligned}
    \op{H}(t)
    &=
    \wL(t) \op{a}_{\mathrm{L}}^{\dagger} \op{a}_{\mathrm{L}}
    +
    \wR(t) \op{a}_{\mathrm{R}}^{\dagger} \op{a}_{\mathrm{R}}
    +
    \wq(t) \op{\sigma}_+ \op{\sigma}_-
    \\
    &\quad
    +
    g_{\mathrm{LR}}(t)
    \left( \op{a}_{\mathrm{L}}^{\dagger} + \op{a}_{\mathrm{L}} \right)
    \left( \op{a}_{\mathrm{R}}^{\dagger} + \op{a}_{\mathrm{R}} \right)
    \\
    &\quad
    -
    \im g_{\mathrm{Rq}}(t)
    \left( \op{a}_{\mathrm{R}}^{\dagger} + \op{a}_{\mathrm{R}} \right)
    \left( \op{\sigma}_+ - \op{\sigma}_- \right),
  \end{aligned}
\end{equation}
where $\op{a}^\dagger_{\mathrm{L}}$, $\op{a}^\dagger_{\mathrm{R}}$ and
$\op{\sigma}_{+}$ are the creation operators for the left oscillator, right
oscillator and qubit, respectively. The first three terms in
Eq.~\eqref{system-hamiltonian} describe the free evolution of the subsystems,
with $\omega_{\mathrm{q}/\mathrm{L}/\mathrm{R}}(t)$ being the time-dependent and
controllable level splittings of the qubit, the left and the right oscillator,
respectively. The fourth and fifth term describe how the right oscillator is
bi-linearly coupled to left oscillator and to the qubit with time-dependent
interaction strengths $g_{\mathrm{LR}}(t)$ and $g_{\mathrm{Rq}}(t)$,
respectively.

The Hamiltonian~\eqref{system-hamiltonian} can be simplified by applying
a rotating-wave approximation, assuming \change{$g_{\mathrm{Rq}}
< g_{\mathrm{LR}}^{0} \ll \wR$}, where $g_{\mathrm{LR}}^{0}$ is the resonant
coupling strength between the oscillators. This results
in~\cite{Tuorila.npjQuantumInf.3.27}
\begin{equation} \label{system-hamiltonian-RWA}
  \begin{aligned}
    \op{H}(t)
    &\simeq
    \wL(t) \op{a}_{\mathrm{L}}^{\dagger} \op{a}_{\mathrm{L}}
    +
    \wR(t) \op{a}_{\mathrm{R}}^{\dagger} \op{a}_{\mathrm{R}}
    +
    \wq(t) \op{\sigma}_+ \op{\sigma}_-
    \\
    &\quad
    +
    g_{\mathrm{LR}}(t)
    \left(
      \op{a}_{\mathrm{L}}^{\dagger} \op{a}_{\mathrm{R}}
      +
      \op{a}_{\mathrm{R}}^{\dagger} \op{a}_{\mathrm{L}}
    \right)
    \\
    &\quad
    +
    \im g_{\mathrm{Rq}}(t)
    \left(
      \op{a}_{\mathrm{R}}^{\dagger} \op{\sigma}_-
      -
      \op{a}_{\mathrm{R}} \op{\sigma}_+
    \right).
  \end{aligned}
\end{equation}
Within this approximation, the number of excitations is a conserved quantity in
the case of unitary evolution. Therefore, the total Hilbert space $\mathcal{H}$
of the system can be conveniently divided into subspaces $\mathcal{H}_{N}$ where
the number of excitations $N$ is constant. A state belonging to a subspace
$\mathcal{H}_{N}$ will thus remain within the subspace during the evolution that
is solely governed by Hamiltonian~\eqref{system-hamiltonian-RWA}.

However, we consider the three-partite system to be open, interacting with an
environment through one of its subsystems. Specifically, we take the left
oscillator to be linearly coupled to a thermal reservoir. \change{Since we want
this coupling to be relatively strong (compared to other typical relaxation
rates), the right oscillator is needed as an intermediate component in order to
allow efficient decoupling of the qubit from the reservoir.} The system-bath
interaction Hamiltonian is of the form~\cite{Tuorila.npjQuantumInf.3.27}
\begin{equation} \label{interaction-hamiltonian}
  \op{H}_{\mathrm{int}}
  =
  \alpha \left(\op{a}_{\mathrm{L}}^{\dagger} + \op{a}_{\mathrm{L}}\right)
  \op{V}_{\mathrm{R}},
\end{equation}
where $\op{V}_{\mathrm{R}}$ is an operator of the reservoir and $\alpha$ plays
the role of an effective coupling strength. In order to derive a master equation
for the open system, we employ its instantaneous eigenbasis
$\{\ket{\Psi_n(t)}\}$, defined by $\op{H}(t) \ket{\Psi_n(t)} = \omega_n(t)
\ket{\Psi_n(t)}$, with $\omega_{n}(t)$ being the respective eigenvalue. In this
representation, the system-bath interaction can be rewritten as
\begin{equation} \label{eq:H_int}
  \op{H}_{\mathrm{int}}
  =
  \alpha \sum_{m,n} v_{mn} \ket{\Psi_m(t)} \bra{\Psi_n(t)} \op{V}_{\mathrm{R}},
\end{equation}
where
\change{%
  \begin{align} \label{eq:v_mn}
    v_{mn}(t)
    =
    \bra{\Psi_m(t)} (\op{a}_{\mathrm{L}}^{\dagger} + \op{a}_{\mathrm{L}})
    \ket{\Psi_n(t)}.
  \end{align}
}
Using standard techniques based on a weak-coupling hypothesis and the Born,
Markov and secular approximations~\cite{breuer2002}, it is possible to derive
a Markovian master equation for the open system. The decay rates, responsible
for dissipation and decoherence, are given by
\begin{equation} \label{eq:Gamma_mn}
  \Gamma_{mn}(t)
  =
  \alpha^2 |v_{mn}(t)|^2 S_{\mathrm{R}} \big[\omega_{mn}\left(t\right)\big],
\end{equation}
where $\omega_{mn}(t) = \omega_m(t) - \omega_n(t)$ and $S_{\mathrm{R}}(\omega)$
is the real part of the Fourier transform of the reservoir correlation function,
\begin{equation} \label{spectral-function}
  S_{\mathrm{R}}(\omega)
  =
  \int_{-\infty}^{+\infty} \mathrm{d}s\, e^{\im \omega s}
  \Braket{\op{V}_{\mathrm{R}}(s) \op{V}_{\mathrm{R}}(0)}_{\mathrm{R}},
\end{equation}
where the average $\braket{\dots}_{\mathrm{R}}$ is taken over the thermal state
of the reservoir \change{and the operators are expressed in the interaction
picture with respect to the bath Hamiltonian}. The corresponding master equation
in the Lindblad form reads
\begin{align} \label{full-master-equation}
  \frac{\dd}{\dd t} \op{\rho}(t)
  &=
  - \im \left[\op{H}(t), \op{\rho}(t)\right]
  \notag \\
  &\quad
  +
  \sum_{m,n} \Gamma_{mn}(t) \Big(%
    \op{L}_{mn}(t) \op{\rho}(t) \op{L}_{mn}^{\dagger}(t)
    \notag \\
    & \qquad \qquad \qquad \quad
    - \frac{1}{2} \big\{%
      \op{L}_{mn}^{\dagger}(t) \op{L}_{mn}(t), \op{\rho}(t)
    \big\}
  \Big),
\end{align}
where the Lindblad operators $\op{L}_{mn}(t) = \ket{\Psi_{m}(t)}
\Bra{\Psi_{n}(t)}$ describe transitions among the eigenstates. \change{A
derivation of the master equation can be found in Appendix~\ref{app}.} The
Hamiltonian $\op{H}(t)$ can be directly controlled by tuning the level
splittings $\omega_{\mathrm{q/L/R}}(t)$. Importantly, the Lindblad operators and
decay rates inherit the temporal dependence from the instantaneous eigenstates
and eigenvalues. As a consequence, Eq.~\eqref{full-master-equation} goes beyond
the description based on static decay channels with constant rates,
\change{although we have neglected the correlations arising from the interplay
between the temporal dependence of the Hamiltonian and the dissipation}.

Solving the full master equation~\eqref{full-master-equation} is a rather
challenging task and we therefore limit our study to a finite number of
subspaces $\mathcal{H}_{N}$. Specifically, we consider the dynamics of the open
system in the two subspaces $\mathcal{H}_{0}$ and $\mathcal{H}_{1}$, i.e., the
subspace with no excitations, $\mathcal{H}_{0} = \mathrm{span}\{\ket{0,0,g}\}$,
and that with a single excitation, $\mathcal{H}_{1}
= \mathrm{span}\{\ket{0,0,e}, \ket{0,1,g}, \ket{1,0,g}\}$ where
$\ket{0,0,g} = \ket{0}_{\mathrm{L}} \otimes \ket{0}_{\mathrm{R}} \otimes \ket{g}_{\mathrm{q}}$,
$\ket{0,0,e} = \ket{0}_{\mathrm{L}} \otimes \ket{0}_{\mathrm{R}} \otimes \ket{e}_{\mathrm{q}}$,
$\ket{0,1,g} = \ket{0}_{\mathrm{L}} \otimes \ket{1}_{\mathrm{R}} \otimes \ket{g}_{\mathrm{q}}$,
$\ket{1,0,g} = \ket{1}_{\mathrm{L}} \otimes \ket{0}_{\mathrm{R}} \otimes \ket{g}_{\mathrm{q}}$.
In the restricted Hilbert space, the Hamiltonian reads
\begin{align} \label{eq:H}
  H(t)
  =
  \begin{pmatrix}
    0 & 0 & 0 & 0
    \\
    0 & \wq(t) & - \im g_{\mathrm{Rq}}(t) & 0
    \\
    0 & \im g_{\mathrm{Rq}}(t) & \wR(t)   & g_{\mathrm{LR}}(t)
    \\
    0 & 0 & g_{\mathrm{LR}}(t) & \wL(t)
  \end{pmatrix}
\end{align}
in the basis $\{\ket{0,0,g}, \ket{0,0,e}, \ket{0,1,g}, \ket{1,0,g}\}$. This
simplified model can be solved analytically in the basis of the instantaneous
eigenstates $\ket{\Psi_{1}(t)}, \ket{\Psi_{2}(t)}, \ket{\Psi_{3}(t)} \in
\mathcal{H}_{1}$ and the ground state $\ket{\Psi_{0}} = \ket{0,0,g}$.

Accounting exclusively for population decay from the excited states in
$\mathcal{H}_{1}$ to the ground state $\ket{0,0,g}$, but not for the reverse
process of thermal excitation~\footnote{\change{%
  For environmental temperatures $T_{\mathrm{env}} \sim \SI{10}{\milli\kelvin}$,
  typical for dilution refrigerators, and qubit frequencies of
  $5-\SI{10}{\giga\hertz}$, typical for superconducting
  qubits~\cite{Gambetta2017}, the thermal occupation of states with double or
  higher excitations is much less than $1\%$. Thermally induced excitation
  processes can thus be neglected~\cite{Tuorila.npjQuantumInf.3.27}. For much
  higher temperatures, in contrast, subspaces with higher excitation numbers
  would become relevant and thermal excitations need to be taken into account.
}},
we obtain the following Lindblad master equation,
cf. Eq.~\eqref{full-master-equation},
\begin{align} \label{eq:LvN}
  \frac{\dd}{\dd t} \op{\rho}(t)
  &=
  \lop{L}(t)\left[\op{\rho}(t)\right]
  =
  - \im \left[\op{H}(t), \op{\rho}(t)\right]
  +
  \lop{L}_{D}(t)\left[\op{\rho}(t)\right]
\end{align}
where
\begin{align} \label{eq:Liouvillian}
  \lop{L}_{D}(t)\left[\op{\rho}(t)\right]
  &=
  \sum_{i=1}^{3} \Gamma_{i0}(t) \Big(%
    \op{L}_{i}(t) \op{\rho}(t) \op{L}_{i}^{\dagger}(t)
    \notag \\
    & \qquad \qquad \qquad
    - \frac{1}{2} \big\{%
      \op{L}_{i}^{\dagger}(t) \op{L}_{i}(t), \op{\rho}(t)
    \big\}
  \Big)
\end{align}
and the three time-dependent Lindblad operators are given by
\begin{align}
  \op{L}_{i}(t)
  =
  \ket{\Psi_{0}}\Bra{\Psi_{i}(t)},
  \qquad
  i=1,2,3.
\end{align}
Closed form expressions for the exact eigenvalues $\omega_{i}(t)$ and
eigenstates $\ket{\Psi_{i}(t)}$, albeit rather lengthy, are straightforward to
calculate with computer algebra.

\change{Note that in addition to the tunable, engineered environment created by
the left oscillator and the resistor, there exists in general also
uncontrollable environments giving rise to the usual background lifetimes. Since
the optimization scheme is essentially independent of such weak background
coupling, we do not consider it further in this work.}

\subsection{Physical realization}
The model introduced above is quite general. In the following, we focus on
a possible experimental realization which implies certain constraints and
specific functional dependencies between the bare frequencies of the three
subsystems, $\{\wL(t), \wR(t), \wq(t)\}$, and their respective couplings. The
model described by Hamiltonian~\eqref{system-hamiltonian} can be realized by
means of a superconducting qubit coupled to two $LC$
resonators~\cite{Tuorila.npjQuantumInf.3.27}. The resonators behave effectively
as quantum harmonic oscillators, the tunable frequencies of which are determined
by the capacitance $C$ and a controllable inductance $L$, i.e.,
$\omega_{L/R}(t)= 1/\sqrt{L_{L/R}(t) C_{L/R}}$. In this implementation, the
couplings between the components can be expressed as functions of the physical
parameters of the system and the bare resonator
frequencies~\cite{Tuorila.npjQuantumInf.3.27}.

The reservoir is realized by connecting a resistor to the left resonator with
$\op{V}_{\mathrm{R}}$ in the interaction
Hamiltonian~\eqref{interaction-hamiltonian} describing voltage fluctuations over
the resistor. The resistor can be modeled as a thermal bath of bosonic
modes~\cite{clerk2010}, with the bath correlation
function~\eqref{spectral-function} corresponding to the Johnson-Nyquist
spectrum,
\begin{equation} \label{Johnson–Nyquist-spectrum}
  S_{\mathrm{R}}(\omega)
  =
  \frac{2 R \omega}{1 - e^{- \omega/k_{\mathrm{B}} T_{\mathrm{env}}}},
\end{equation}
where $R$ is the resistance of the resistor and $T_{\mathrm{env}}$ denotes its
electron temperature~\cite{clerk2010}. At low temperature, the spectral
function~\eqref{Johnson–Nyquist-spectrum} strongly suppresses emission of
thermal excitations from the resistor so that indeed the population decay is the
leading-order dissipative process for the studied three-partite quantum system.
The decay rates can be expressed as~\cite{Tuorila.npjQuantumInf.3.27}
\begin{align} \label{eq:rates}
  \Gamma_{i0}(t)
  &=
  \Gamma_{0} \left|\Braket{%
    \Psi_{0}(t) | \op{a}_{\mathrm{L}}^{\dagger} + \op{a}_{\mathrm{L}}
    | \Psi_{i}(t)
  }\right|^{2}
  \notag \\
  &\quad \times
  \frac{\wL(t) \omega_{i}(t)}{\wR(t)^{2}}
  \frac{1}{1 - e^{- \omega_{i}(t)/k_{\mathrm{B}} T_{\mathrm{env}}}},
\end{align}
where $\Gamma_0$ plays the role of a static decay rate. \change{Note that the
decay rates fulfill the detailed balance
condition~\cite{Tuorila.npjQuantumInf.3.27}
\begin{align} \label{eq:db}
  \Gamma_{mn}(t)
  =
  \mathrm{exp}\left\{%
    - \frac{\omega_{mn}(t)}{k_{\mathrm{B}} T_{\mathrm{env}}}
  \right\}
  \Gamma_{nm}(t),
\end{align}
which implies suppression of thermal excitations at low temperatures.
}

\subsection{Quantum Optimal Control Theory}
In general, QOCT aims at finding the optimal external control fields to steer
the dynamics of a quantum system in the desired way~\cite{Glaser.EPJD.69.279}.
The starting point is to express the optimization task
as a 
functional of
the yet unknown external control fields $\{\pulse_{k}(t)\}$,
\begin{equation} \label{eq:J_oct}
  \begin{aligned}
    J\left[\left\{\pulse_{k}\right\}\right]
    &=
    \alpha_{\tau}\left[\left\{\op{\rho}_{l}(\tau)\right\}\right]
    \\
    &\quad+
    \int_{0}^{\tau} \dd t\, g\left[\left\{\pulse_{k}(t)\right\},
    \left\{\op{\rho}_{l}(t)\right\}, t\right]\,.
  \end{aligned}
\end{equation}
\change{Here, $\alpha_{\tau}$ denotes the final-time target functional which
describes the actual optimization task such as the preparation of
a specific target state. It} may depend on one or several states
$\op{\rho}_{l}(\tau)$, \change{where the subscript $l$ denotes the different
initial conditions of the temporal evolution}. Furthermore, $g$ describes
constraints that are relevant also at
intermediate times, such as constraints on the intensity or spectrum \change{of
the yet unknown control fields}~\cite{PalaoPRA13, ReichJMO14}. A proper choice
of the functional \change{requires} that the extremum is attained \change{if and
only if the task is carried out in an optimal way. In the example of state
preparation, this is the case if the system state matches the desired target
state perfectly.} \change{In the following, we discuss the two terms in the
optimization functional~\eqref{eq:J_oct} in more detail.}

The final-time functional $\alpha_T$ \change{measures} how well the target is
reached.  For the reset task at hand, we seek to prepare the qubit in its ground
state, irrespective of the initial state of the total system. This can be
achieved by considering the dynamics for several initial states
$\{\op{\rho}_{l}(t=0)\}$, making sure that all of them result in the desired
target state~\cite{ReichNJP13}. Moreover, no excitation should be left in any of
the two oscillators, since otherwise these might get transferred to the qubit in
an uncontrolled fashion later on. Hence, our set of initial states
$\{\op{\rho}_{l}(t=0)\}$, is given by any complete basis of the excited subspace
$\mathcal{H}_{1}$. The respective target state is the ground state of the total
system $\op{\rho}_{\mathrm{trg}} = \ket{\Psi_{0}} \bra{\Psi_{0}}$. The
final-time functional reads
\begin{align} \label{eq:alpha_T}
  \alpha_{\tau}
  =
  1 - \frac{1}{3} \sum_{l=1}^{3} \left\langle
  \op{\rho}_{\mathrm{trg}}, \dmap(\tau,0;\{\pulse_{k}\}) \op{\rho}_{l}
  \right\rangle,
\end{align}
where $\langle \op{A}, \op{B} \rangle = \mathrm{Tr}\{\op{A}^{\dagger} \op{B}\}$
and $\dmap(\tau,0;\{\pulse_{k}\})$ is the control-dependent dynamical map.
\change{Since $\alpha_{\tau}$ measures the remaining population in
$\mathcal{H}_{1}$ at final time $\tau$,} it corresponds to the error of the
reset protocol. An ideal protocol is given by $\alpha_{\tau} = 0$,
\change{%
  which can be attained if and only if no population is left in
  $\mathcal{H}_{1}$. The final-time functional $\alpha_{\tau}$ provides
  a measure for how far the final state of the system is away from the desired
  target. It does not contain any information about the dynamics that brought it
  there.
}

\change{%
  Although our aim is to minimize Eq.~\eqref{eq:alpha_T}, which quantifies the
  reset error, we will achieve this by minimization of the total
  functional~\eqref{eq:J_oct}. To this end,
}
we employ Krotov's method~\cite{Krotov.book, Konnov.AutomRemContr.60.1427}, an
iterative optimization algorithm that comes with the advantage of monotonic
convergence.
\change{%
  Note that in Krotov's method, the function $g$ is needed even if we do not
  want to impose constraints on the control fields or the system dynamics. In
  particular, the choice of $g$ determines the update rule for the control
  fields $\{\pulse_{k}(t)\}$~\cite{PalaoPRA03, Reich.JCP.136.104103}.
}
Given the final-time target, a choice of $g$ in Eq.~\eqref{eq:J_oct}, and the
equation of motion for the system, Eq.~\eqref{eq:LvN}, Krotov's method provides
a recipe to derive an \change{optimization algorithm to determine}
$\pulse_{k}$~\cite{Reich.JCP.136.104103}. Here, we use the standard choice of
minimal amplitude increase per iteration step~\cite{PalaoPRA03},
\begin{align} \label{eq:krotov:g}
  g\left[\left\{\pulse_{k}(t)\right\}\right]
  =
  \sum_{k} \frac{\lambda_{k}}{S_{k}(t)} \bigg[
    \pulse_{k}(t) - \pulse_{k}^{\mathrm{ref}}(t)
  \bigg]^{2},
\end{align}
where $\pulse_{k}^{\mathrm{ref}}(t)$ is a reference field for each
$\pulse_{k}(t)$, taken to be the field from the previous iteration, $S_{k}(t)
\in (0,1]$ a shape function to smoothly switch the field modulations on and off,
and $\lambda_{k}$ a parameter that controls the update magnitude of
$\pulse_{k}(t)$ in each optimization step.
\change{%
  Due to the choice of $\pulse_{k}^{\mathrm{ref}}(t)$ to be the control field
  from the last iteration, the difference $\pulse_{k}(t)
  - \pulse_{k}^{\mathrm{ref}}(t)$ approaches zero as the optimization converges.
  Hence, the contribution of $g$ to the total functional~\eqref{eq:J_oct}
  decreases as well. As the optimum is approached, the value of the overall
  functional~\eqref{eq:J_oct} is essentially given by the value of
  $\alpha_{\tau}$ as desired.
}

With Eq.~\eqref{eq:krotov:g}, the update equation for $\pulse_{k}(t)$
reads~\cite{Reich.JCP.136.104103}
\begin{widetext}
  \begin{align} \label{eq:krotov:update}
    \pulse_{k}^{(i+1)}(t)
    =
    \pulse_{k}^{(i)}(t)
    +
    \frac{S_{k}(t)}{\lambda_{k}} \ip\left\{%
      \sum_{l} \Braket{%
        \op{\chi}^{(i)}_{l}(t)\;,\;
        \frac{\partial \lop{L}\left[\left\{\pulse_{k'}\right\}\right]}{\partial
        \pulse_{k}} \Big|_{\{\pulse^{(i+1)}_{k'}(t)\}}
        \op{\rho}^{(i+1)}_{l}(t)
      }
    \right\},
  \end{align}
\end{widetext}
where $\{\op{\rho}^{(i+1)}_{l}(t)\}$ are the forward propagated initial states
spanning $\mathcal{H}_{1}$, obtained by solving
\begin{equation} \label{eq:krotov:fw_states}
  \frac{\dd}{\dd t} \op{\rho}^{(i+1)}_{l}(t)
  =
  - \im \lop{L}\left[\left\{\pulse^{(i+1)}_{k}\right\}\right]
  \op{\rho}^{(i+1)}_{l}(t)\,.
\end{equation}
The so-called co-states $\{\op{\chi}^{(i)}_{l}(t)\}$ in
Eq.~\eqref{eq:krotov:update} are solutions of the adjoint equation of motion,
\begin{equation} \label{eq:krotov:co_states}
  \frac{\dd}{\dd t} \op{\chi}^{(i)}_{l}(t)
  =
  \im
  \lop{L}^{\dagger}\left[\left\{\pulse^{(i)}_{k}\right\}\right]
  \op{\chi}^{(i)}_{l}(t)\,,
\end{equation}
with boundary condition $\op{\chi}^{(i)}_{l}(\tau)
= - \nabla_{\op{\rho}_{l}(\tau)} \alpha_{\tau}
\big|_{\{\op{\rho}^{(i)}_{l'}(\tau)\}}$. The derivative of $\alpha_{\tau}$ with
respect to $\op{\rho}_{l}(\tau)$ can be turned into a usual gradient by
representing the states in a complete, orthonormal basis~\cite{PalaoPRA03}. Note
that the indices $(i+1)$ and $(i)$ indicate values for current and last
iteration, respectively.

As with any optimization algorithm based on variational calculus, Krotov's
method requires the calculation of gradients---one of them the gradient of the
dynamical generator with respect to the controls, cf.
Eq.~\eqref{eq:krotov:update}. Peculiarly, not just the Hamiltonian $\op{H}$, cf.
Eq.~\eqref{eq:H}, but also the dissipator $\lop{L}_{D}$, cf.
Eq.~\eqref{eq:Liouvillian} depends on the controls and thus contributes to the
gradient,
\begin{align}
   \frac{\partial \lop{L}\left[\left\{\pulse_{k'}\right\}\right]}{\partial
   \pulse_{k}} \op{\rho}
   =
  - \im
  \left[\frac{\partial \op{H}\left[\left\{\pulse_{k'}\right\}\right]}{\partial
  \pulse_{k}}, \op{\rho}\right]
  +
  \frac{\partial \lop{L}_{D}\left[\left\{\pulse_{k'}\right\}\right]}{\partial
  \pulse_{k}} \op{\rho}.
\end{align}
Whereas the gradient of the Hamiltonian with respect to $\wL$, $\wR$ and $\wq$
is straightforward to calculate, cf. Eq.~\eqref{eq:H}, the gradient of the
dissipator $\lop{L}_{D}$ is rather lengthy to evaluate, cf.
Eq.~\eqref{eq:Liouvillian}. This inconvenience is due to the dependence of the
decay rates $\Gamma_{i0}$ and Lindblad operators $\op{L}_{i}$ on the
instantaneous eigenvalues $\omega_{i}(t)$ and eigenstates $\ket{\psi_{i}(t)}$.
The required derivatives of $\omega_{i}(t)$ and $\ket{\psi_{i}(t)}$ with respect
to $\wL$, $\wR$, and $\wq$ have been algebraically calculated using computer
software.

\section{Numerical Results}\label{sec:results}

\begin{figure}[tb]
  \centering
  \includegraphics{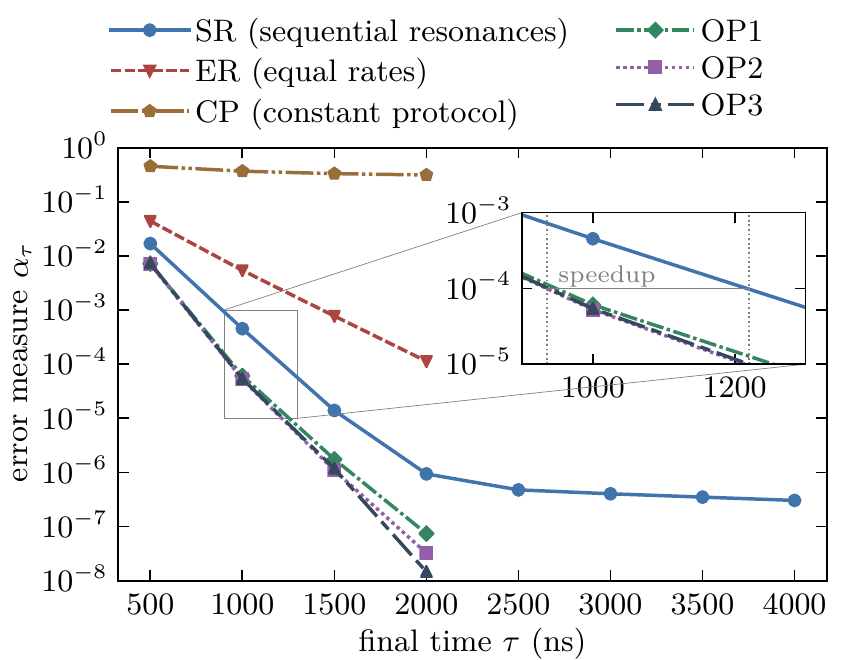}
  \caption{%
    Excited state population $\alpha_{\tau}$, Eq.~\eqref{eq:alpha_T}, as
    a function of protocol length $\tau$ for different control fields.
    SR denotes the original protocol utilizing sequential resonances with the
    resistive bath~\cite{Tuorila.npjQuantumInf.3.27}, CP refers to a protocol
    with only constant fields, and OP1, OP2, and OP3 are results obtained with
    SR or CP \change{as guess control fields} to start the optimization
    \change{(see main text for detailed explanations)}.
    An optimization targeting equal dissipation rates, cf. Eq.~\eqref{eq:Ri},
    instead of minimizing $\alpha_{\tau}$ is labeled by ER.
    The inset highlights the speedup due to the optimization, by comparing the
    durations for which the optimized protocols and the SR reach an error of
    $10^{-4}$.
    The parameters are summarized in Table~\ref{tab:params}.
  }
  \label{fig:JT}
\end{figure}

\begin{table}
  \centering
  \caption{%
    Parameters used in the calculations for the setup shown in
    Fig.~\eqref{fig:scheme}.
    \change{%
      The parameters are taken from Ref.~\cite{Tuorila.npjQuantumInf.3.27} and
      correspond to an experimentally feasible circuit QED realization. Here,
      $T_{\mathrm{env}}$ is a typical temperature for dilution refrigerators
      used to operate superconducting qubits~\cite{Gambetta2017}.
    }
  }
  \begin{tabular*}{\linewidth}{l@{\extracolsep{\fill}}cr}
    \hline
    left oscillator frequency  & $\wLz/2\pi$                & $11.5$\,GHz   \\
    right oscillator frequency & $\wR/2\pi$                 & $10.0$\,GHz   \\
    qubit frequency            & $\wq/2\pi$                 & $9.5$\,GHz    \\
    right osc.-qubit coupling  & $g_{\mathrm{Rq}}/2\pi$     & $68$\,MHz     \\
    left-right osc.\ coupling  & $g_{\mathrm{LR}_{0}}/2\pi$ & $74$\,MHz     \\
    static decay rate          & $\Gamma_{0}$               & $31$\,MHz     \\
    temperature                & $T_{\mathrm{env}}$         & $10$\,mK\,\,\,\\
    \hline
  \end{tabular*}
  \label{tab:params}
\end{table}

\subsection{Optimization of the original protocol}
The original protocol~\cite{Tuorila.npjQuantumInf.3.27} is based on a simple
choice for the left-oscillator frequency $\wL(t)$. Effectively, it consists of
two stages, namely $\wL(t) = \omega_{+}$ and $\wL(t) = \omega_{-}$, separated by
an intermediate ramp. The entire protocol reads
\begin{equation} \label{eq:omega_L}
  \wL(t)
  =
  \left\{%
    \begin{alignedat}{3}
      & \wLz \rightarrow \omega_{+},
      && \qquad 0 && \leq t < t_{\mathrm{R}};
      \\
      & \omega_{+},
      && \qquad t_{\mathrm{R}} && \leq t < \tau/2;
      \\
      & \omega_{+} \rightarrow \omega_{-},
      && \qquad \tau/2 && \leq t < \tau/2+t_{\mathrm{R}};
      \\
      & \omega_{-},
      && \qquad \tau/2+t_{\mathrm{R}} && \leq t < \tau-t_{\mathrm{R}};
      \\
      & \omega_{-} \rightarrow \wLz,
      && \qquad \tau-t_{\mathrm{R}} && \leq t \leq \tau;
    \end{alignedat}
  \right.
\end{equation}
where $\tau$ is the total protocol duration, $\tau/2$ is the hold time at each
stage, and $t_{\mathrm{R}}\ll\tau$ is the ramping duration. \change{The ramp
formula has been chosen to be
\begin{align}
  \wL(t)
  =
  \omega_{0} + \left(\omega_{1} - \omega_{0}\right)
  f\left(\frac{t - t_{0}}{t_{1} - t_{0}}\right)
\end{align}
with $f(x) = 6 x^{5} - 15 x^{4} + 10 x^{3}$, which ramps $\wL(t)$ smoothly from
$\omega_{0}$ to $\omega_{1}$ as time goes from $t_{0}$ to $t_{1}$.
}
The operation points $\omega_{+}$ and $\omega_{-}$ have been chosen such that
$\Gamma_{20}(t) = \Gamma_{30}(t)$ in the case of $\omega_{+}$ and
$\Gamma_{10}(t) = \Gamma_{20}(t)$ for $\omega_{-}$. This choice guarantees that
any excitation decays at some point of time during the protocol.

Figure~\ref{fig:JT} illustrates the performance of the protocol of sequential
resonances (SR) with the resistive bath
as a function of its duration $\tau$. It shows a rapid approach towards errors
$\alpha_{\tau}$ as small as $10^{-6}$ for $\tau=\SI{2000}{\nano\second}$ for the
parameters listed in Table~\ref{tab:params}. Although this may be sufficient for
some applications, the SR exhibits a plateau for longer durations, preventing it
even theoretically to reach significantly smaller errors. The plateau is caused
by population being locked in the excited state of the right oscillator --- an
unfavorable feature that is apparently not resolvable by simply extending the
protocol duration. However, taking Eq.~\eqref{eq:omega_L} as the initial guess
for the above described optimization procedure, Fig.~\ref{fig:JT} shows that,
depending on $\tau$, an improvement of up to two orders of magnitude in the
error $\alpha_{\tau}$ compared to the SR is possible. In addition, this
optimized protocol (OP1) also resolves the issue of the plateau, reaching errors
$\alpha_{\tau} < 10^{-7}$. The improvement with respect to the protocol duration
is comparatively modest, as the inset of Fig.~\ref{fig:JT} illustrates. Taking,
e.g., $\alpha_{\tau}=10^{-4}$ as a sufficiently small error, the speedup with
respect to the SR is roughly $\Delta \tau \approx \SI{280}{\nano\second}$.

\begin{figure}[tb]
  \centering
  \includegraphics{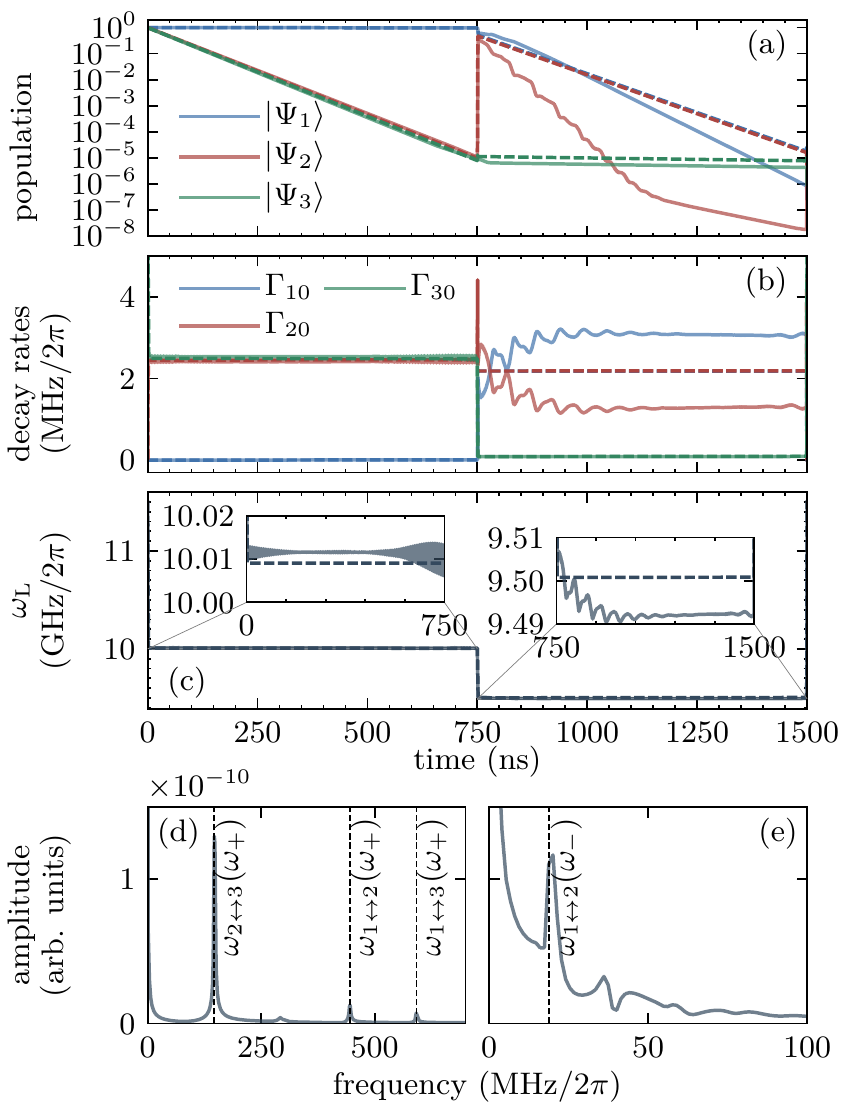}
  \caption{%
    Dynamics for the SR (dashed lines) and its optimized version OP1 (solid
    lines) for a protocol duration of $\tau=\SI{1500}{\nano\second}$, cf.
    Fig.~\ref{fig:JT}.
    \change{Note that dashed lines partly overlap.}
    (a) Population in the three eigenstates of the excited subspace
    $\mathcal{H}_{1}$.
    (b) Decay rates, cf. Eq.~\eqref{eq:rates}, from $\mathcal{H}_{1}$ into the
    total ground state $\ket{\Psi_{0}}$.
    (c) Left oscillator frequency $\wL(t)$ following the original two stage
    protocol of Eq.~\eqref{eq:omega_L}. The two stages are still visible in
    the optimized version, with modulations on top, as highlighted by the two
    insets. The shaded area in the left inset corresponds to fast oscillations,
    which are not resolved due to the linewidth.
    (d) and (e) show the frequency spectra of the optimized splitting $\wL(t)$
    from the left and right insets of (c), respectively. The vertical lines
    indicate frequency differences, $\omega_{i \leftrightarrow j}
    = \left|\omega_{i} - \omega_{j}\right|$ with $\omega_{i} = \omega_{i}(\wL)$
    being the instantaneous eigenvalues.
  }
  \label{fig:L}
\end{figure}

Figure~\ref{fig:L} compares the decay dynamics of SR and OP1, for
$\tau=\SI{1500}{\nano\second}$, showing the population of the excited
eigenstates in Figs.~\ref{fig:L}(a) and the respective decay rates and control
fields $\wL(t)$ generating them in Fig.~\ref{fig:L}(b) and \ref{fig:L}(c). We
observe that the original two-stage protocol (SR) acts as intended, i.e., the
population decays from all three eigenstates of $\mathcal{H}_{1}$. Since the
intermediate ramp transfers a significant amount of population from
$\ket{\Psi_{1}}$ to $\ket{\Psi_{2}}$, $\Gamma_{20}(t)$ needs to be sufficiently
large also during the second stage. Note that this population transfer between
different eigenstates within $\mathcal{H}_{1}$ occurs due to non-adiabatic
transitions caused by changes of those particular
eigenstates~\footnote{\change{Note that the population transfer between
different eigenstates due to non-adiabatic transitions is accompanied by
a contribution originating from the change in the eigenstates themselves}}.
These are caused by changes in the control function $\wL(t)$, i.e., the ramps in
the SR.

A similar reasoning readily explains also the \change{behavior of the control
field in case of} OP1, shown in Fig.~\ref{fig:L}(c). Compared to the SR, the
optimization effectively shifts the base levels of \change{$\wL$ at} both stages
and adds oscillations on top. This results in an increase of $\Gamma_{10}(t)$
and a decrease of $\Gamma_{20}(t)$, cf.\ Fig.~\ref{fig:L}(b), in particular
during the second stage, directly causing the population of $\ket{\Psi_{1}(t)}$
($\ket{\Psi_{2}(t)}$) to decay faster (slower). The additional oscillations,
even though having small amplitude, drive non-adiabatic transitions between
$\ket{\Psi_{1}}$ and $\ket{\Psi_{2}}$, which primarily transfer population to
the fast decaying state $\ket{\Psi_{1}}$, cf.\ Fig.~\ref{fig:L}(a). This becomes
even more clear by inspecting Figs.~\ref{fig:L}(d) and \ref{fig:L}(e), which
show the spectra corresponding to the insets of Fig.~\ref{fig:L}(c). In both
cases, the frequencies match the differences between various eigenvalues
$\omega_{i}$, evaluated at $\omega_{+}$ and $\omega_{-}$ for $\wL$ in the left
and right inset, respectively. Whereas the spectrum shown in Fig.~\ref{fig:L}(d)
is dominated by a peak at $\omega_{2 \leftrightarrow 3}$, which does not seem to
have a notable impact on the dynamics, Fig.~\ref{fig:L}(e) exhibits a peak at
$\omega_{1 \leftrightarrow 2}$ and is responsible for the above-mentioned
population transfer between $\ket{\Psi_{1}}$ and $\ket{\Psi_{2}}$. The
combination of increasing decay rates and engineered population transfer results
in the excitation to more efficiently decay from both states. \change{The
required control of the left oscillator frequency $\wL(t)$ can, for instance, be
achieved by Josephson parametric amplifiers~\cite{APL.93.042510}.}

The optimization studied in Fig.~\ref{fig:L} changes the coherent part of the
evolution compared to the SR, creating non-adiabatic transitions by suitably
modulating $\wL(t)$ and adapting the decay rates $\Gamma_{i0}(t)$ accordingly.
Both effects are necessary to explain the observed improvement with respect to
the SR\@. In contrast, Fig.~\ref{fig:JT} shows also
optimization results where the system dynamics has been completely ignored in
the optimization process. In this case, the minimization of
Eq.~\eqref{eq:J_oct} has been replaced by a functional targeting equal
dissipation rates (ER). Namely, we have optimized $\wL(t)$ to yield $R_{1}
\approx R_{2} \approx R_{3}$ with each $R_{i}$ as large as possible, where
\begin{align} \label{eq:Ri}
  R_{i} = \int_{0}^{\tau} \Gamma_{i0}(t) \,\dd t,
  \qquad
  i = 1,2,3,
\end{align}
are the time-integrated dissipation rates which are independent of the system
dynamics. The naive assumption behind this optimization is that, since all
states $\op{\rho}_{1}, \op{\rho}_{2}, \op{\rho}_{3}$ are equally weighted in
Eq.~\eqref{eq:alpha_T}, equal dissipation from all of them may be a good
choice to decrease the error $\alpha_{\tau}$. However, this is not
the case, cf.\ Fig.~\ref{fig:JT}, which emphasizes the interplay of coherent and
dissipative dynamics in the problem at hand.

\subsection{Optimization with an extended set of control fields}

\begin{figure}[tb]
  \centering
  \includegraphics{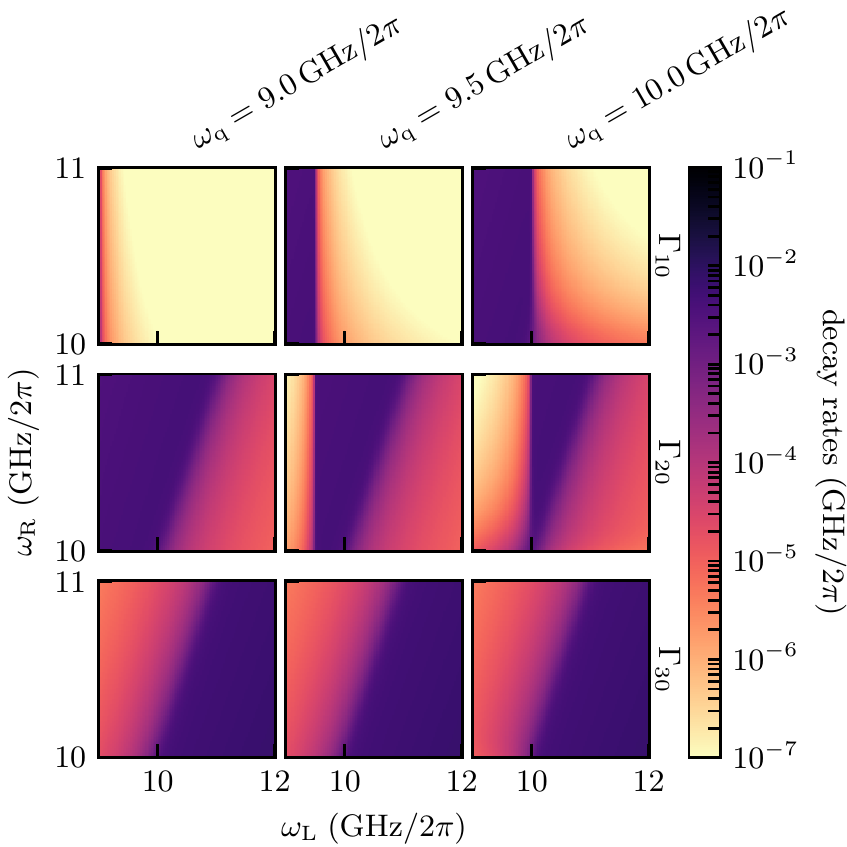}
  \caption{%
    Decay rates $\Gamma_{i0}$ from the excited subspace $\mathcal{H}_{1}$ into
    the total ground state $\ket{\Psi_{0}}$, cf.\ Eq.~\eqref{eq:rates}, as
    a function of level splittings $\omega_{\mathrm{L}}$ and
    $\omega_{\mathrm{R}}$ and for three different values of
    $\omega_{\mathrm{q}}$.
  }
  \label{fig:rates_map}
\end{figure}

In the following, we extend the SR by assuming the frequencies of the right
oscillator and of the qubit, $\wR(t)$ and $\wq(t)$, to be temporally
controllable. Since the eigenvalues $\omega_{i}(t)$ and eigenstates
$\ket{\Psi_{i}(t)}$ ($i=1,2,3$) depend on all three frequencies, $\wL$, $\wR$
and $\wq$, changing any of them may affect the dynamics. In other words, more
control fields give the optimization more flexibility to steer the system
dynamics in the desired way and engineer the dissipation rates more
appropriately.

\begin{figure}[tb]
  \centering
  \includegraphics{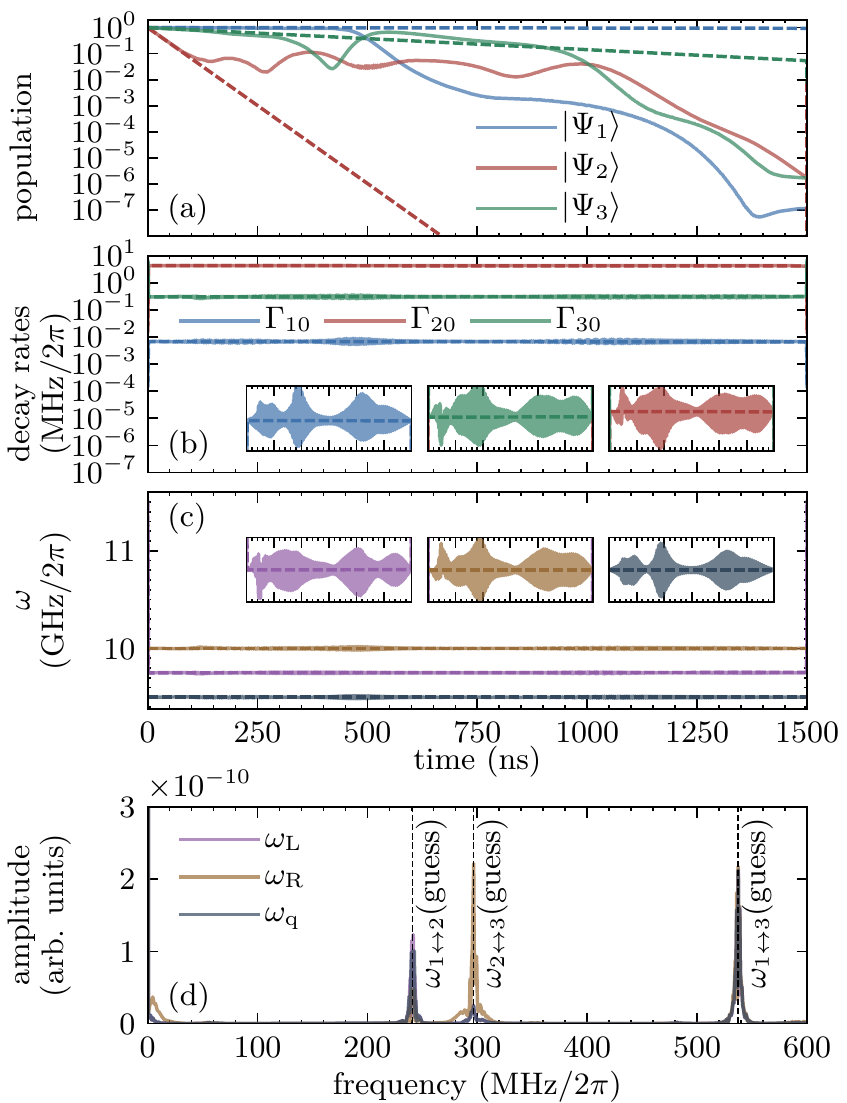}
  \caption{%
    Dynamics obtained with the constant protocol CP (dashed lines) and its
    optimized version OP3 (solid lines).
    The panels are as in Fig.~\ref{fig:L} with the small insets in (b) and (c)
    providing a closer look at the shapes of the optimized fields, respectively
    decay rates, compared to their non-optimized, constant counterparts.
    Panel (d) shows the spectra of all optimized fields from panel (c).
  }
  \label{fig:LRq}
\end{figure}

First, we inspect in Fig.~\ref{fig:rates_map} how the decay rates change as
a function of the level splittings $\wL$, $\wR$, and $\wq$. Two important
observations can be made from Fig.~\ref{fig:rates_map}. On one hand, the decay
rates are still mutually exclusive, in the sense that there exists no
combination such that two of them are \change{maximal} at the same time. On the
other hand, the attainable \change{total} maximum \change{of each individual
decay rate as a function of all three controls $\wL$, $\wR$ and $\wq$} does not
change. Hence, adjusting $\wR$ or $\wq$ in addition to $\wL$ does not yield
essentially larger rates, and there will not be a significantly faster decay to
the ground state. Although no \change{naive} improvement is to be expected
\change{from simply increasing the} decay rates, i.e., due to the dissipative
part of the dynamics, one may still achieve an improvement by more appropriately
steering the coherent part.

Figure~\ref{fig:JT} shows optimization results for the case that all three
frequencies are time-dependent (OP2). The initial guess has been chosen
according to the SR, i.e., Eq.~\eqref{eq:omega_L} for $\wL(t)$ and constant
values for $\wR$, $\wq$. Despite the extended set of controls, the optimization
does not yield errors significantly below the case where only $\wL(t)$ is
controlled. This finding is reproducible even when using different sets of
controls, such as only using $\wq(t)$ and $\wL$ or only using $\wq(t)$ and $\wR$
(data not shown). We therefore expect that further controls beyond $\wL(t)$ do
not allow the coherent part of the dynamics to be steered more efficiently.

In order to study this expectation further and evaluate the impact of the guess
fields, we have carried out optimizations with all three possible controls.
Whereas $\wR$ and $\wq$ have been set constant as initial guess, $\wL(t)$ has
been chosen as $\wL(t) = (\omega_{+} + \omega_{-})/2$ with additional ramps in
the beginning and end. Due to this choice, $\Gamma_{20}$ is almost maximal
during the entire protocol, whereas $\Gamma_{10}$ and $\Gamma_{30}$ are orders
of magnitude smaller, cf.\ Fig.~\ref{fig:rates_map}. Thus, only population in
$\ket{\Psi_{2}}$ decays fast. Simply extending the protocol duration $\tau$
will not solve the problem of small $\Gamma_{10}$ and $\Gamma_{30}$. Upon
optimization, we are, however, able to find fields yielding similarly small
errors $\alpha_{\tau}$ as before, \change{cf. OP3 with OP1 and OP2 in
Fig.~\ref{fig:JT}.} We again analyze an
exemplary dynamics for $\tau=\SI{1500}{\nano\second}$ in Fig.~\ref{fig:LRq}.
Figure~\ref{fig:LRq}(a) shows the population dynamics. As expected, the
population in $\ket{\Psi_{2}}$ decays rapidly under the constant guess fields,
while $\ket{\Psi_{3}}$ exhibits only slow decay and the population in
$\ket{\Psi_{1}}$ is almost conserved. The respective decay rates and control
fields are shown in Figs.~\ref{fig:LRq}(b) and \ref{fig:LRq}(c). Interestingly,
the optimization leaves the base levels of each control field unchanged, again
adding small oscillations on top. Consequently, the decay rates are unchanged in
magnitude but exhibit small oscillations as well. Since $\Gamma_{20}$ is already
maximal by choice of the guess fields, cf.\ Fig.~\ref{fig:rates_map}, there is
no possibility for the optimization to increase it. Instead, the optimization
ensures that all excitations are coherently transferred to this strongly
decaying state --- in our example from $\ket{\Psi_{1}}$ and $\ket{\Psi_{3}}$ to
$\ket{\Psi_{2}}$, as evident from Fig.~\ref{fig:LRq}(a). Thus, we find a similar
reset strategy as in Fig.~\ref{fig:L}: The control fields are tailored such that
a single decay rate (not necessarily the same at different times) is maximal and
population is transferred coherently into this strongly decaying state.

We expect the reset strategies illustrated in Figs.~\ref{fig:L}
and~\ref{fig:LRq} to be feasible for essentially any combination of control
fields and choice of guess fields. This follows from the decay rates being
mutually exclusive, cf. Fig.~\ref{fig:rates_map}, i.e., \change{if one state has
a maximal decay rate, the other two states decay slower.} All that is hence
required is to ensure coherent population transfer into this state which seems
to be possible by tailoring the control fields. Remarkably, the addition of
further control fields does not result in significantly smaller errors
$\alpha_{\tau}$, cf.\ Fig.~\ref{fig:JT}. In fact, $\wL(t)$ alone is already
sufficient to fully control the decay rates and engineer the required population
transfer. Nevertheless, adding more control options increases flexibility and is
thus potentially beneficial in experiments, especially if certain control fields
are convenient to implement experimentally.

\section{Summary and Conclusions} \label{sec:concl}
In summary, we have studied how optimization of external control fields speeds
up the initialization of a superconducting qubit which is tunably coupled to
a thermal bath via two resonators. The control knobs are the time-dependent
level splittings of the qubit and the resonators. Starting from a protocol
utilizing sequential resonances with the resistive bath and employing the level
splitting of a single resonator as the only control field, while assuming the
initial state to be confined to the single excitation
subspace~\cite{Tuorila.npjQuantumInf.3.27}, we have replaced the analytically
derived temporal dependence by a numerically optimized control field. This has
allowed us to obtain an improvement in both the reset speed and fidelity.

We have also tested whether adding multiple control fields, by explicitly
accounting for the tunability of the level splitting of the qubit and of the
second resonator, results in additional improvements. This has turned out to not
to be the case. Moreover, we have found that in all control scenarios, the
optimized reset strategy consists in maximizing the decay rate from a single
state and driving non-adiabatic population transfer into the strongly decaying
state by small oscillations in the control fields. Even for different
combinations of control fields and various guess fields, the optimization has
resulted in reset errors and times of the same order of magnitude. We thus
suspect to have identified the quantum speed limit for qubit reset in this
particular physical setup with tunable couplings, provided that only a single
excitation \change{at maximum} is present initially. However, a more rigorous
study exploring the full parameter space is required to prove that our solution
represents indeed a global, and not only a local, optimum.

Whether the quantum speed limit identified in our study is related to
the rotating-wave or other used approximations remains an open question. In
particular, it will be interesting to study whether the reset duration and error
can be further decreased by utilizing couplings between the single-excitation
subspace and higher-excitation subspaces. The rationale would be that highly
excited states decay faster which might further decrease the protocol duration.
The required transitions could again be driven by suitably shaped control fields
determined by QOCT.

Our study is, to the best of our knowledge, the first demonstration of
experimentally directly applicable reservoir engineering using quantum optimal
control of time-dependent decay rates. It is related to earlier results obtained
for controlling open quantum systems with non-Markovian dynamics which had
shown, for example, improved cooling due to cooperative effects of control and
dissipation~\cite{SchmidtPRL11} or better gate operations~\cite{RebentrostPRL09,
ReichSciRep15}. Our approach differs from the more common scenario for the
control of open quantum systems in which the external field modifies only the
Hamiltonian and thus the coherent part of the dynamics, rather than the
dissipator of the master equation~\cite{KochJPCM16}. In contrast, in our
example, both the coherent evolution and the decay rates change
\change{in time} as a result of the field
optimization~\footnote{\change{Note that our approach differs from approaches
like dynamical decoupling, where the decay rates are effectively modified by
control fields but actually remain time-independent.}}. Specifically, the
changes in the coherent dynamics are manifested in the occurrence of
non-adiabatic transitions which go hand in hand with modifications in the
time-dependent decay rates. Interestingly, coherent and dissipative dynamics
are tightly intertwined and the optimization protocol affects both in
a physically transparent way. Our study thus paves the way to explore quantum
reservoir engineering in condensed phase settings.

\begin{acknowledgments}
  Financial support from the Volkswagenstiftung, the DAAD, the Academy of
  Finland via the QTF Centre of Excellence program (projects 287750, 312058,
  312298, and 312300) is gratefully acknowledged. This research was supported in
  part by the National Science Foundation under Grant No. NSF PHY-1748958 and
  the European Research Council Grant No. 681311 (QUESS).
\end{acknowledgments}

\begin{appendix}
  \section{Derivation of the Master Equation} \label{app}
  In this appendix, we provide details on how to obtain the Lindblad master
  equation~\eqref{full-master-equation}. It follows in large parts the
  derivation in Ref.~\cite{Tuorila.npjQuantumInf.3.27}. We know that the
  combined dynamics of the system and the environment follows is unitary and
  obeys the von Neumann equation
  \begin{align}
    \frac{\dd}{\dd t} \op{\rho}_{\mathrm{tot}}(t)
    &=
    - \im \left[\op{H}_{\mathrm{tot}}(t), \op{\rho}_{\mathrm{tot}}(t)\right],
  \end{align}
  where $\op{\rho}_{\mathrm{tot}}(t)$ is the joint state of the system and the
  environment and
  \begin{align}
    \op{H}_{\mathrm{tot}}(t)
    =
    \op{H}(t) + \op{H}_{\mathrm{env}} + \op{H}_{\mathrm{int}},
  \end{align}
  is the total Hamiltonian, $\op{H}_{\mathrm{env}}$ is the Hamiltonian of the
  environment alone and $\op{H}(t)$ and $\op{H}_{\mathrm{int}}$ are given by
  Eqs.~\eqref{system-hamiltonian-RWA} and~\eqref{interaction-hamiltonian},
  respectively. In order to obtain an equation of motion for the reduced
  dynamics of the system alone, i.e., $\op{\rho}(t) = \mathrm{tr}_{\mathrm{env}}
  \left\{\op{\rho}_{\mathrm{tot}}(t)\right\}$, we start with applying a unitary
  transformation $\op{D}(t) = \sum_{n} \Ket{\psi_{n}} \Bra{n}$ that diagonalizes
  $\op{H}(t)$, where $\{\ket{n}\}$ is a time-independent basis. In the new basis
  of eigenstates $\{\ket{\Psi_{n}(t)}\}$ of $\op{H}(t)$, the system Hamiltonian
  reads
  \begin{align} \label{eq:app:H_eff}
    \op{H}_{\mathrm{eff}}(t)
    &=
    \sum_{n,m} \left[
      \omega_{n}(t) \delta_{n,m} - \im \Braket{\Psi_{n}(t) | \dot{\Psi}_{m}(t)}
    \right]
    \ket{n}\bra{m},
  \end{align}
  where $\omega_{n}(t)$ is the corresponding eigenvalue of $\ket{\Psi_{n}(t)}$.
  The second term in Eq.~\eqref{eq:app:H_eff} is responsible for non-adiabatic
  couplings between different eigenstates. The derivation of the master equation
  starts with conventional assumptions like initial separability,
  $\op{\rho}_{\mathrm{tot}}(0) = \op{\rho}(0) \otimes
  \op{\rho}_{\mathrm{env}}(0)$, a thermal and static state of the bath
  $\op{\rho}_{\mathrm{env}}(t) \approx \op{\rho}_{\mathrm{env}}(0)$, weak
  coupling between the system and its environment and the typical Born-Markov
  and secular approximations~\cite{breuer2002}. We obtain the general Lindblad
  master equation~\eqref{full-master-equation},
  where the decay rates $\Gamma_{mn}(t)$ are still undefined. However, it can be
  shown that the decay rates coincide with the ones that can be obtained with
  Fermi's golden rule~\cite{Alicki1977}. This yields the general expression in
  Eq.~\eqref{eq:Gamma_mn} for the decay rates. By substituting $v_{mn}(t)$ and
  $S_{\mathrm{R}}$ from Eqs.~\eqref{eq:v_mn}
  and~\eqref{Johnson–Nyquist-spectrum}, respectively, one arrives at the decay
  rates in Eq.~\eqref{eq:rates}. Due to the form of the system-environment
  interaction~\eqref{eq:H_int}, dephasing processed described by rates
  $\Gamma_{nn}(t)$ vanish. Moreover, since the detailed balance~\eqref{eq:db}
  holds and taking the temperature of the environment low, heating processes are
  strongly suppressed and cooling is the dominant source of dissipation.

  Note that the frame, where the Hamiltonian $\op{H}(t)$ is diagonal, i.e.,
  Eq.~\eqref{eq:app:H_eff}, is only used for the derivation of the decay rates.
  In the numerical simulations, all operators and states are still expressed in
  the static basis $\{\ket{0,0,g}, \ket{0,0,e}, \ket{0,1,g}, \ket{1,0,g}\}$.
\end{appendix}


%

\end{document}